\documentclass[11pt]{article}
\usepackage{moriond,epsfig}

\def\be{\begin{equation}}
\def\ee{\end{equation}}
\def\bea{\begin{eqnarray}}
\def\eea{\end{eqnarray}}

\begin{document}
\vspace*{4cm}
\title{RECENT RESULTS FROM CHORUS CHARM ANALYSIS}

\author{  M. G\"{u}ler \\
(for the CHORUS collaboration)}

\address{METU,  06531 Ankara, Turkey}

\maketitle\abstracts{
The CHORUS experiment was designed to search for $\nu_{\mu}\rightarrow \nu_{\tau}$ oscillation by detecting the decay topology of
 the $\tau$ in charged current (CC) $\nu_{\tau}$ events.
 The detector was exposed to the Wide Band Neutrino Beam of the
 CERN SPS during the years 1994-1997.  About $10^{6} \nu_{\mu}$ CC
 events were collected in the nuclear emulsion target.  Up to now,
 about 170,000 $\nu_{\mu}$ events have been located and analysed.
 The speed of the automated emulsion scanning systems increases each
 year.  With the present performance of these systems, it has become
 possible to perform large volume scanning.  All tracks belonging
 to an interaction vertex can be recognized and measured precisely.
 This technique is not only  applied to the search for neutrino
 oscillation but can also be used for the recognition of events where
 charmed particles are produced.
 Results obtained from the analysis of a sub-sample of the data on the
 production rate in $\nu_\mu$ CC interactions of neutral charmed mesons
 ($D^0$) and charmed baryons ($\Lambda_c$) are presented.  In
 addition a new measurement of the branching ratio for  the 
decays of 
 charmed hadrons into  muons is given.
 Also measurements of topological branching ratios of $D^0$ and
 $\Lambda_c$ are presented. Finally, a search for associated  charm production 
is discussed.}

\section{Introduction}
\subsection{The CHORUS experiment}\label{subsec:prod}

The  CHORUS experiment was proposed primarily to search for
$\nu_{\mu}\rightarrow\nu_\tau$ oscillations through the appearance  of
$\nu_\tau$ in a $\nu_\mu$ beam, aiming to explore the domain of small
mixing angles down to $\sin$2$\theta_{\mu\tau}\sim 3\times10^{-4}$ for
mass parameters
$\Delta$$m^2$$\sim$ 1 e$V^2$. This represents an order of magnitude
improvement over the previous generation of experiments\cite{e531b}.

The CHORUS detector  is a hybrid
set-up that combines nuclear emulsion stacks with various electronic 
detectors.
The nuclear emulsion acts both as target for neutrino interactions and as
detector, allowing three-dimensional reconstruction of short-lived
particles as the    $\tau^-$ lepton and charmed hadrons. The nuclear
emulsion target, which is segmented into four stacks, has an overall mass
of 770 kg, each of the stacks consisting of eight modules of 36 plates
of
size 36$\times$72 cm$^2$. Each plate has a 90
$\mu$m
 plastic support coated on both sides
with  350 $\mu$m emulsion layers. Each stack is followed
by three interface
emulsion sheets having a 90 $\mu$m emulsion layer  on both sides of an
800 $\mu$m thick plastic base  and by a set of scintillating~ fiber~
tracker planes.
The interface sheets and the  fiber trackers provide accurate
particle trajectory predictions into the  emulsion stack in order to
locate the vertex positions.  The accuracy of the fiber tracker prediction
is about
150 $\mu$m in position and  2 mrad in the track angle. The major drawback
is the absence of the time information: any charged particle traversing
the emulsion during the exposure will leave a track.

The emulsion scanning is performed by fully automatic
microscopes  equipped with CCD cameras and a read out system, called
${\it Track~ Selector}$\cite{nakano}. In order
to recognize track segments in an emulsion, a series of tomographic images
are taken by focusing at different depths in the emulsion thickness.  
The digitized images are shifted according to the predicted track angle
and then added.
The presence of  aligned grains forming a track is detected
as a local peak of the gray level of the summed image.
The track finding efficiency of the
track selector is higher than 98$\%$ for track slopes less than 400 mrad.

The electronic detectors downstream of the emulsion target include a
hadron spectrometer
which measures the bending of charged particles in an air-core magnet, a
calorimeter
where the energy and direction of showers are measured and a muon
spectrometer which measures the charge and momentum of muons.
\subsection{Data Collection}
The West Area Neutrino Facility (WANF) at CERN provides  a beam of 27 GeV
average energy
consisting mainly of $\nu_\mu$ with a  5$\%$ $\bar{\nu}_{\mu}$
contamination. For the four years  in which the emulsion target was exposed,
the integrated beam intensity  corresponds  to 5.06$\times 10^{19}$ protons on target. 
The analysis of the data from the electronic detectors allows the identification
of the set of events possibly orginating from the emulsion stacks. For the  
first 
phase of the analysis, the events were subdivided into two classes,  based 
on 
electronic detectors,
which are the  one-muon and the zero-muon samples, distinguished by the 
presence 
or absence of one reconstructed muon 
of negative charge. For  vertex location, track trajectories belonging  to reconstructed 
events in the electronic detectors are used  to guide  the scanning. The  track is first searched 
in the interface sheets and then followed back  into the target emulsion. 
If it is not found in two 
consecutive  plates, the first of these  is defined as {\it  vertex plate}.  This  plate 
may contain  the primary vertex or the decay vertex, or both.  Once 
the vertex 
is 
located, the charm decay search
is performed using the {\it  netscan} method \cite{net}. It consists of recording all track segments 
within an angular acceptance in a volume surrounding the assumed vertex position.
\section{Charm analysis}

Charm production in  neutrino charged current (CC) interactions has
been
studied   in several experiments 
\cite{cdhs}
with electronic detectors and mostly
through the analysis
of dimuon events. In these events, the leading muon is interpreted as
originating from the
neutrino vertex and the other, of opposite charge, as the product of the
charmed particle semileptonic decay.
Experiments of this type, however, suffer from a significant  background
of events ($\sim$ 30$\%$)
 in which  the second muon  originates from an undetected decay
 in flight of a pion or a kaon rather than from a charm decay.
 Moreover, the type of the charmed particle and its decay
topology can not be identified in these experiments.
 Nevertheless they have provided  measurements of the strange  
 quark content of the nucleon as well as an estimation of the charm
 quark mass. A much lower level of background can be achieved using
an emulsion target which provides a sub-micron spatial resolution, and
hence, the
topological identification of charmed hadron decays. The statistics
accumulated
in this way is however
very limited.  Only recently, the development within CHORUS of automatic
scanning
devices of much higher speed  has made studies of
charm production  with high statistics possible.
\subsection{$D^0$ production rate measurement}

In the  past, $D^0$ production  in neutrino interactions was studied 
in E531
experiment\cite{e531b}. Only 57 $D^0$ events  were identified  and analysed. 
In CHORUS, we have performed  an analysis with much higher statistics.
About 25,000 one-muon events  were analysed representing   $\sim$15$\%$ of the CHORUS data.
The main criteria
for the event selection are the
following:  at least one of the decay daughters and muon track
reconstructed with  more than one-segment, matched tracks in the electronic
detector; and 
the daughter track must have a  significant impact parameter with respect 
 to the vertex  point.
These criteria   select  851 events from  25,000 CC
$\nu_\mu$ events. The selected events are visually inspected in order to distinguish decays from hadronic interactions,
gamma conversion and false vertices which are  reconstructed using 
background tracks.
After performing visual inspection, we confirmed 
that out  of 851  selected events,
226 show 2-prong decay topology and 57 show 4-prong decay
topology.
 
The efficiency
of the charm selection was evaluated with a GEANT3 based simulation of the
experiment. Large samples of deep-inelastic neutrino interactions
were generated according to the beam spectrum using the JETTA generator
derived from LEPTO and JETSET. The simulated response of the CHORUS
electronic detectors is processed through the same reconstruction program 
used for data.  To evaluate the {\it netscan} efficiency, realistic conditions
of track densities need to be reproduced. This is achieved by merging the
emulsion data of the simulated events with real netscan data which do not
have a reconstructed vertex but contain tracks which stop or pass through
fiducial  volume representing the real background\cite{sim}. The combined
data are passed through the same netscan reconstruction and selection
programs as used for data. The selection efficiencies are estimated as
(58.6$\pm$5.1$)\%$ and (70.1$\pm$5.2$)\%$ for 2-prong and 4-prong decays respectively.
Based upon 282  $D^0$ decays with an estimated background of 9.2$\pm$2.6  
$K^0$ and $\Lambda^0$, we obtain\cite{d0}:
$\frac{\sigma (D^{0} \rightarrow V2+V4)}{\sigma
(\mathrm{CC})}$=(1.99$\pm$0.13({\it
stat.})$\pm$0.14 ({\it syst.}))$\times 10^{-2}$ at 27
GeV.  The topological ratio V4/V2 is found to be 
\bea
(23.1\pm 4.0)\% 
\eea
Combining  (1) with $Br(D^0\rightarrow V4) = (13.3\pm 07)\times 10^{-2}$
using the  PDG tables\cite{pdg}, we can obtain branching ratio of $D^0$
decaying into  neutral particles:
\bea
Br(D^0\rightarrow neutral) = 1-Br(D^0\rightarrow V4)\times (1+(\frac{Br(D^0\rightarrow V4)}{Br(D^0\rightarrow V2)})^{-1})
\eea
\bea
= (29.1\pm 10.4)\times 10^{-2}\nonumber
\eea
The precision of this measurement will be improved with the  final statistics.

\subsection{Semi-leptonic branching rate  measurement}

To measure the semi-leptonic branching ratio, the event 
 the event 
selection criteria
are the following: at least two tracks of more than one-segment coming 
from different vertices
match tracks at the electronic detectors.
About 50,000 $\nu_\mu$ CC
events are analysed with this selection and in total, 1055 events are
selected.
To estimate the  selection purity, 25$\%$ of selected events are visually
inspected. The selection purity is obtained as 91$\%$. Based on this
selection purity, the number of charmed  hadrons is 956$\pm$35.

The efficiency
of  charm selection was evaluated as explained in the previous section.
The correction factor ({\it R}) for the efficiency  in the determination 
of the
semileptonic branching fraction is defined as
$\frac{\Sigma_{D_{i}}\epsilon_{D_i}f_{D_i}}{\Sigma_{D_{i}}\epsilon_{D^{\mu}_i}f_{D_i}}$
where $\epsilon_{D_i}$ is the selection efficiency for charm species $D_i$
, $\epsilon_{D^{\mu}_i}$ the selection efficiency for semileptonic decays
of  $D_i$, $f_{D_i}$ the fragmentation fraction. Based on the estimated
efficiencies R is  1.01$\pm$ 0.05.

The average semi-leptonic branching fraction can be  written in terms of
measurable quantities as
$B_{\mu}= \frac{N_{2\mu}^{sel.}}{N^{sel.}}\times R$
where $N^{sel.}$  is the number of selected events, corrected
for the selection purity and  $N_{2\mu}^{sel.}$ is the number of selected
events with a secondary muon in the final state corrected for selection
purity as well as the muon identification efficiency and purity.

Combining
these numbers, we measured $B_\mu = 0.0093\pm (stat.)\pm 0.009
(syst.)$.\cite{semi}
\subsection{$\Lambda_c$ production rate  measurement}
Although evidence for charmed baryon production by neutrinos has been
reported in the literature with few events observed in a number of
different  bubble chamber experiments, cross-section values are known
with a large error.

Our analysis is based on a statistical approach using the flight length
distributions of charmed hadrons. Since the lifetime of the $\Lambda_c$   
is smaller than that of other charged charm hadrons, the sample of  
charged  
charm decaying at distances  less than 400 $\mu$m from the $\nu$ vertex, 
called short decays, should be dominated by $\Lambda_c$. Conversely,
long decays are enriched by $D^+$ and $D_{s}^+$ decays. Therefore,
the analysis is performed applying two different selections. Event selection
for short flight decays is based on the following: muon track and
daughter track must be reconstructed  having more than one segment and
the daughter track must have a big impact parameter with respect  to the 
vertex point
(5$\mu$m $<$ IP $<$ 30 $\mu$m). We have analysed about 50,000 CC $\nu$  
events 
with short decay selection criteria. In total, 1614 events were selected 
for visual
inspection. 62 events are  confirmed as 1-prong decay and 66 events show
3-prong decay topology.

On the other hand, for long flight decays we
require that the parent track and muon track must be reconstructed by
the reconstruction  software and
the parent track must have a small impact parameter with respect to the 
neutrino vertex and
at least one of the daughter tracks must be reconstructed
with more than  one segment and minimum distance between parent track and
 daughter track must be less than 5 $\mu$m.
About 56,000 CC $\nu$ events were analysed
and 586 events were  visually inspected. The confirmed charm events after
flight length cut are  the following: 133 1-prong and 195 3-prong decays.

Combining short and long decay search and taking into account efficiency
and background, the number of $\Lambda_c$ is
861$\pm$198(stat.)$\pm$98$(syst.)^{+140}_{-54}(QE)$. Based on these events 
the branching ratio into 3-prongs was determined:
BR$(\Lambda_c\rightarrow 3~ prong)$=0.24$\pm$0.07(stat)$\pm$0.04(syst.)
and the total charged-current cross section\cite{lambda}:
$\frac{\sigma (\Lambda_c)}{\sigma(\mathrm{CC})}$=(1.54$\pm$0.35({\it
stat.})$\pm$0.18 ({\it syst.}))$\times 10^{-2}$.

\subsection{Associated charm production}

Associated charm (c$\bar{c}$) production in neutrino interactions is a 
very  rare process
and therefore  very difficult to observe. In the past, c$\bar{c}$  
production in charged-current 
$\nu_\mu$ interactions has been  studied through trimuons and like-sign dimuons , however, the reported production 
rate is higher than theoretical expectation.  On the hand, only one event consistent with neutral-current (NC) production 
of  c$\bar{c}$ has been observed by the E531 collaboration\cite{e531b}. 
Based on one event, the production rate  
with respect to NC 
production is estimated as $\frac{\sigma (c\bar{c})}{\sigma(\mathrm{NC})}$=(0.13$\pm$0.31) $\times 10^{-2}$. The sub-micron 
resolution of 
nuclear 
emulsion allows the  study of this kind of rare processes. We have 
performed  a search in CC $\nu_\mu$ interactions
 and one event is observed with the characteristics  of  c$\bar{c}$ production\cite{ccbar}.  A new search has been started,  both  in
the NC and CC event sample.
Fig. 1. shows the schematic view of  one of
the candidate events.  This event is  found in the zero-muon sample and  shows two V2 decay topologies.
One minimum-ionizing particle  and three
heavy ionizing particles are emitted from primary  vertex. A neutral  particle($C^0$) decays in the same emulsion  plate,
62 $\mu$m downstream of the primary vertex. In the next plate, another neutral decay is found with the  flight length of
977 $\mu$m. The lifetimes of the decaying particles are estimated using the relation:  $\tau$=$\frac{\it l <\theta>}{\it c}$ 
\cite{life}
where $<\theta>$  is average emission angle of daughter particles  relative to the parent direction , {\it l} is the flight lenght of the 
decaying particle and {\it c} is the speed of light. The estimated lifetimes, shown in table 1, are consistent with the decay of the
neutral charmed particles.
The search for new candidates 
and more detailed analysis  of already observed  events are in progress.
\begin{table}[h]
\caption{List of particles at primary and secondary vertices.}
\vspace{0.4cm}
\begin{center}
\begin{tabular}{|c|c|c|c|l|}
\hline
~	 	&$\theta_{y}(rad)$ 
		&$\theta_{z}(rad)$
		&${\it l}$($\mu$m)   
		&$\tau$(s)\\ \hline
particle1	&-0.060
		&-0.002
		&~
		&~\\ \hline
$1-{C^0}$	&0.008
		&0.031
		&63
		&$2.6\times 10^{-14}$\\\hline
daughter (at 1-${C^0}$)	&-0.019
			&0.040	
			&~
			&~\\\hline
daughter (at 1-${C^0}$)	&-0.052
			&0.028
			&~
			&~\\\hline
2-${C^0}$		&-0.078
			&0.095
			&977 
			&$5.5\times 10^{-13}$\\\hline
daughter (at 2-${C^0}$)	&-0.166
			&0.049
			&~
			&~\\\hline
daughter (at 2-${C^0}$)	&0.035
			&0.103	
			&~
			&~\\\hline
\end{tabular}
\end{center}
\end{table}

\begin{center}
\begin{figure}
\epsfig{figure=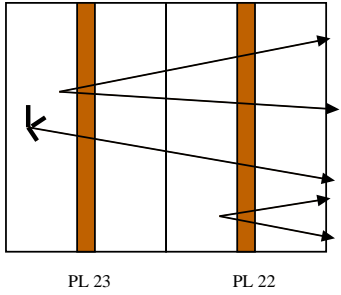 ,height=1.5in, angle=0}
\caption{Sketch of the event topology in emulsion. 
\label{fig:radish}}
\end{figure}
\end{center}

\end{document}